\documentstyle[12pt,epsf]{article}
\def\be{\begin{equation}}
\def\ee{\end{equation}}
\def\bea{\begin{eqnarray}}
\def\eea{\end{eqnarray}}
\def\ba*{\begin{eqnarray*}}
\def\ea*{\end{eqnarray*}}
\begin{document}

\newcommand{\sheptitle}
{Thoughts on Defining the Multiverse }
\newcommand{\shepauthor}
{ Laura Mersini-Houghton}
\newcommand{\shepaddress}
{ Department of Physics and Astronomy, UNC-Chapel Hill, NC, 27599-3255, USA}

\begin{titlepage}
\begin{flushright}
%hep-ph/yymmnnn\\
%SNS-PH/02-04\\
\today
\end{flushright}
\vspace{.1in}
\begin{center}
{\large{\bf \sheptitle}}
\bigskip \medskip \\ \shepauthor \\ \mbox{} \\ {\it \shepaddress} \\
\vspace{.5in}
%{\bf Abstract}

\bigskip \end{center} \setcounter{page}{0}

\newcommand{\shepabstract}
\shepabstract

%\baselineskip 24pt

%\maketitle

%%%%%%%%

{The possibility that the multiverse corresponds to physical reality deserves serious investigation. Having three different important theories,(quantum mechanics, string theory and inflation), predict the existence of the multiverse is hardly coincidental. I argue that the existence of the multiverse must be expected from the underlying fundamental theory or else we can not meaningfully address the puzzle of the initial conditions. In this view, the extension of our current cosmology to a multiverse framework becomes an extension of the Copernican principle to nature. In order to discuss the ontology of the multiverse I propose to apply:\\
- the principle of 'No Perpetual Motion' as a criterion for the parameter of time; and,\\
- the principle of 'Domains Correlations' as a criterion for determining the background spacetime in which the multiverse is embedded.} 
%\begin{abstract}
%\end{abstract}
%{\Large \bf Equation of state }

\begin{flushleft}
\hspace*{0.9cm} \begin{tabular}{l} \\ \hline {\small Email:
mersini@physics.unc.edu} \\

\end{tabular}
\end{flushleft}

\end{titlepage}
%%%%%%%%%%%%%%%%%%%%%%%%%%%%%%%%%%%%%%%%%%%%%%%%%%%%%%%%%%%%%%%%%%%%%%%%%%%%%%%%%%%%%

\section{Multiverse Theories}

Currently three of our fundamental and most successful theories predict the existence of a multiverse. They are Quantum Mechanics, String Theory and the theory of Inflation, briefly described below. Currently the multiverse field is experiencing a renewal of interest and investigation. As I have argued recently, this may be due to the fact that the emergence of a multiverse from theories of nature is not a coincidence. Rather, predicting a multiverse may prove to become a requirement for the theory of quantum gravity. The motivation for defining the multiverse, along with the argument for needing to extend our conceptual framework to include the multiverse are discussed in Sec.2. In Sec.3, I conjecture the application of two principles in the multiverse that could help us define and investigate important questions for its ontology as well as provide a handle for testable predictions. Known theoretical examples of multiverses at present are: The Everett, Eternal Inflation and String Theory Landscape multiverses.

\subsection{ Everett Multiverse}
In the late 1950's Hugh Everett III gave his many worlds interpretation of Quantum Mechanics according to which, in the family of wavefunctions obtained from solving the Schrodinger equation, each wavefunction corresponds to a perfectly good physical reality (world). The Everett interpretation of Quantum Mechanics was complemented in the 1970's by Zeh\cite{zeh} and Zurek\cite{zurek} who showed that the emergence of a classical world from the quantum wavefunction is achieved via a mechanism known as decoherence. Since then, the latter has been tested experimentally. The Everett multiverse, although debated for over 50 years, did provide the first scientific example of a multiverse being predicted from an underlying physical theory.

\subsection{ Eternal Inflation Multiverse}

The paradigm of inflation has been tremendously successful in explaining many of the observed features in our universe, such as flatness, homogeneity, structure and the scale-invariance of the cosmic microwave background (CMB) as well as being in nearly perfect agreement with the current observations so far \cite{wmap}. It should be noted however that inflation comes with a heavy price tag by creating a set of new puzzles, such as: why our universe started with such improbable initial conditions ; why was the initial patch so smoothly fine-tuned ; and, what is the inflaton? Some of these puzzles may appear less severe in the framework of eternal inflation. Current thinking has it that in many of the inflation models, once inflation starts it never entirely switches off, i.e. that there are domains within our horizon that randomly end up with a large enough fluctuation to start inflating, thereby producing new bubble universes\cite{guth}. If our universe continually reproduces new ones then the ensemble of all these bubbles provides a multiverse predicted by eternal inflation. The problem of the probability distribution of these bubbles in the eternal multiverse, known as the measure problem, remains an open issue. The weight assigned to the likelihood of having such large fluctuations and therefore eternal inflation, have been recently challenged and debated\cite{lauraparker,aguirre}.

\subsection{ String Landscape Multiverse}

Advances in String Theory over the last five years led to the discovery of the landscape of string theory\cite{landscape,ashok,douglas,bousso}. String theorists discovered that after reducing from higher dimensions to $(3+1)-D$ worlds, they did not end up with one unique solution but rather something like $10^{500}$ vacua solutions, the energy profile of which comprises the landscape. Work along this direction continues and it is possible that the number of string vacua solutions on the landscape, that contain $(3+1)-D$ worlds like our universe, will grow. In 2005, I proposed to view the ensemble of landscape vacua solutions, as the string theory multiverse \cite{laurareview, land1} and argued that, since every vacua on the landscape is a potential birth place for a universe like ours, then the ensemble of these vacua is to be interpreted as a multiverse prediction of string theory\cite{laurareview, land1,land2,laurarich1}. An equivalent interpretation of the above proposal is to consider the string theory multiverse during its quantum phase, to be the physical phase space for the initial conditions\cite{land1,land2,laurarich1,laurarich2}. By proposing to place the wavefunction of the universe on the landscape and use its many-worlds interpretation\cite{land1,land2,laurareview}, the quantum mechanics multiverse was thus embedded onto the string landscape, thereby making the $Everett$ and $Landscape$ multiverses equivalent\cite{laurareview,laurarich1,laurarich2}.

\section{ Why do we need to live in a Multiverse?}

Breakthroughs in observational astrophysics have led us to an intriguing picture of the universe by challenging our previous understanding and expectations of nature.

Ten years ago we discovered that our universe is accelerating again, just like during the Big Bang inflation, but at lower energies. The acceleration is attributed to the most mysterious form of energy, dark energy. Since then, many other anomalies in the CMB have been spotted at the largest scales, scales compared to the horizon size. Such a lack of understanding of the present universe thus pushes the mystery of the selection of the initial conditions and inflation into the forefront of research. 

Naturally we expect that a deeper understanding of the mysteries of the birth of our universe and of its current acceleration, would stem from the theory of quantum gravity. But breakthroughs in the leading candidate for quantum gravity, string theory, led to a discovery of the landscape that is still considered by many to be a bizarre picture. The general view was that the multitude of vacua discovered presents a problem since we had expected that string theory would yield one unique solution corresponding to a universe that: started with high energy inflation; contained the correct values for the constants of nature; and, contained a tiny but nonzero amount of dark energy tuned exquisitely to $122-$ orders of magnitude, which was just right for dominating the expansion of the universe at recent times. But string theory does not predict a unique universe, on the contrary, it predicts a multiverse$!$

A radically different point of view was advocated in \cite{laurareview,laurarich1,laurarich2}, namely: if we are ( even allowed) to ask these deep questions about the selection of our universe then a multiverse picture must be an expected prediction for the theory of quantum gravity. How else can we meaningfully ask: why did we start with such an extremely improbable universe without implying as compared to what other possibilites?  Basically, a statement about an extraordinarily improbable event carries meaning only if we accept that other more probable events can be conceived. In this context then, the problem of the selection of the initial conditions implies that an extension of our theories to the multiverse framework is neither a philosophical contemplation, nor an unfortunate coincidence among our cherished theories, but could be a physical reality that is required in order to gain a deeper understanding of nature.

In fact, the extension of our current cosmology to a multiverse framework is an extension of the Copernician Principle to nature. Over the ages we found that we are not at the center of our solar system, that the solar system is not at the center of the universe and now, that the universe is not at the center of the world (the multiverse).

\section{What is the Multiverse: How Many Types are there?}

 If nature provided us with a multiverse, allowing for a discussion of the ontology of the multiverse and the potential for extracting testable predictions, becomes of central importance in cosmology. Let us assume that the emergence of the background spacetime(s) has already been addressed by the underlying theory of quantum gravity, which also predicts the multiverse. Often in literature \cite{weinberg,davies,tegmark} the term 'multiverse' is informally used to denote any multitude of universes. But by now we need a detailed discussion of issues such as their spacetime background, the relation between the different species, and ultimately their observational imprints. As described in Sec.1 we have already come across three seemingly unrelated examples for a multiverse. They can not all correspond to the same physical reality unless identical, because if they are distinguishable than we have simply transfered the puzzle of the selection of the initial conditions for our universe to a new puzzle, the selection of the 'real' multiverse. The main question thus becomes: what is the multiverse, how many distinguishable species are there and, in what spacetime does the multiverse reside?

%%%%%%%%%%
Let us define the {\it 'multiverse'} to be the ensemble of all possible universes predicted by the underlying theory. For the sake of definiteness in terminology, consider the{\it ' universe'} to be (conservatively) defined as the domain of spacetime in which points were causally connected at some time slice, say $t=0$, of their past light cones. In an influential paper in 2003 \cite{tegmark} Tegmark classified the multiverses into four levels, (see also \cite{rudy}). Since then, as already discussed in Sec.1., advances in string theory and precision cosmology have provided a wealth of information and, possibly observational imprints\cite{rudnick, avatars1,avatars2} relevant to the multiverse. The discussion here about the questions listed above on the number of distinguishable multiverse species, their background space and parameter of time can be viewed as an analysis of a multiverse corresponding to Tegmark's level $4$, (since his levels $1,2,3$ are subsets of level $4$), which is inclusive of the predictions made by the theories described in Sec.1 for the ensembles of universes comprising a multiverse.
By now it is important that we provide a set of rules and an agreed taxonomy for the emerging field of multiverse physics. For simplicity let us discuss separately the different species of multiverses comprised from $(3+1)-D$ universes, and the types based on ensembles of universes with varying dimensionality. Within the multiverse, each domain can have:\\
A) its own set of laws ;\\ 
B) its own set of constants of nature; or,\\
C) its own dimensionality\\
that vary from one domain to the next across the multiverse.

If our notion of a quantum to classical transition holds generically for some of the categories above, then just like the universe members, we can further fine-split the above multiverse species to two phases of existence, the quantum phase and the classical phase. The quantum multiverse is the phase space of the initial conditions for all members of the ensemble living in configuration space, whereas the classical multiverse is the ensemble of all the classical universes, embedded in a spacetime background, which are born from the quantum phase after the quantum to classical transition occurs, (perhaps similar to  the phenomenon of decoherence for our universe). It is possible that at some stage a mixed phase of quantum and classical universes may co-exist in the multiverse. I will focus only on the pure phases before the transition switches on and after the quantum to classical transition has been completed throughout the multiverse, and discuss below in a bit more detail the amount of information that can be extracted for each of the species ($A, B, C$), and in particular, the issue of the background space and time. 

%%%%%%%%%%

\subsection{ (Type A): Different laws across the Multiverse:} 

In the type $A$ multiverse, by definition if we use the set of laws and equations that comprise our physical theories, we can not make predictions about the other domains since their physical laws are unknown to us and our theories are not relevant to them. These domains are beyond the realm of our understanding at present. 

What is time in the multiverse? Do we have the gravitational force in common with the other domains? Can gravitons from our domain leak across into the other sectors of the multiverse?  Many such issues remain open at present. Even if the answers to the latter question is yes which implies potential for observational imprints in our universe, we would still lack the computational power for making those predictions. It is possible that the multiverse predicted  by eternal inflation may fall into the type $A$ category in which case the severe difficulties with the measure problem may be conceptual rather than technical.

\subsection{ (Type B): Same laws but different constants of nature across the Multiverse:}

In the type $B$ multiverse, there definitely exists a force of gravity in all the domains since the laws and equations are the same as ours across this multiverse. In type $B$ gravity can spread everywhere and thus correlations among different domains do exist. In type $B$ multiverses we have the potential to make predictions about domains beyond our causal horizon. The Everett multiverse and the string theory landscape multiverse may be two such example of the type $B$ multiverses, since in both cases there is an overall embedding theory from which the ($3+1$)-D members of the ensemble (solutions) are derived, and which are closely related to each-other\cite{laurareview,laurarich1,land1,land2}.

\subsection{ (Type C ): Varying dimensionality Multiverse :} This multiverse is the ensemble of universes comprised from domains have different dimensions. In general, any subset of Type $C$ that contain $(N+1)-D$ universes, ($N>3$), can be further be branched in the two groups $A^{N}$ and $B^{N}$ where $A^N, B^N$ correspond to the types $A, B$ defined above but in space dimensions $N>3$. This division is possible only if the parameter of time is fundamental in the type$ C$ multiverse. This argument makes it plausible for Type $A$ and Type $B$ multiverses to be sectors in the parent multiverse of Type $C$ since they could correspond to the sector of $C$ with $N=3$, where the higher $N$ dimensional branches are stacked above. 

Clearly a knowledge of what is time in the $C$ multiverse is crucial for its understanding and its relevance to our sector $N=3$.
Yet, since currently we lack a deeper understanding of this challenging issue, below I would like to propose that we apply the 'Principle of No Perpetum Mobile' for all multiverse species in order to achieve consistency in our guesstimates for their basic characteristics and embbedding spacetime.

\subsection{ Is Time in the Multiverse Fundamental or Emergent?\\   The Principle of 'No Perpetual Motion'}

Understanding the nature of time is another central mystery in theoretical physics. We have seen in Sec.2 that the question: is time fundamental or emergent, becomes crucially important to the multiverse physics. 

In Type $A$ multiverses it is possible to allow for time to emerge since the laws of physics vary across the multiverse. Specifically we can think of a Schrodinger equation which relates the parameter of time to the energy level of the wavefunction. If this equation is not valid for the whole ensemble then it is not difficult to envision a scenario where the parameter of time appears differently in various domains since it emerges from their different corresponding theories. We could also imagine that time acquires different meanings in different domains. In such a scenario it is plausible that the parameter of time may not appear as a fundamental quantity. This question becomes more severe for the type $C$ multiverse. 

In Type $B$ multiverses, since the laws of nature are the same everywhere, then the definition of the time parameter would also be the same. In this type, the existence of time as a fundamental property of nature appears to be the more plausible scenario.

Although there is no agreement yet on the issue of time being emergent or fundamental\cite{timearrow}, it is clear that
the nature of time and the multiverse are closely inter-related. Understanding one of these fundamental questions will shed light on the other and vice-versa.
Even though these problems remain open, based on the known symmetries and basic categorization of the multiverses, we can still derive some general statements about the parameter of time in the multiverse if we make the assumption that quantum mechanics is not an effective but a universal theory. If time translation is a symmetry of the multiverse (type $A,B$ or $C$) and the time-energy relation holds, then energy conservation resulting from this symmetry, implies that {\it 'no perpetual motion'} is allowed across the multiverse. At this stage of our knowledge we can not derive this conclusion since we dont know if the relation between time translation symmetry and energy conservation holds. 

Therefore I would like to propose that we impose the {\it The Principle of No Perpetual Motion in the Multiverse}, which requires that energy can not be created or destroyed in or across domains in the multiverse. A by-product of this principle is to reduce the number of multiverse species that are candidates for physical reality to only those where the parameter of time does not vary across the multiverse, and is likely to be fundamental. This principle could reduce the number of possible multiverses corresponding to physical reality, to only those where the parameter of time is the same for all the domains, and possibly to narrow down the candidates to solely the multiverse where time appears fundamental.

\section{ In what spacetime does the Multiverse exists? \\  The Principle of 'Domains Correlations'}

Given that only one of the multiverses described above corresponds to the physical reality, and assuming that, independent of the space dimensionality, based on the principle of 'no perpetual motion', there is only one time parameter in that multiverse, then 
a crucial question is: {\it In what background spacetime does the multiverse exist?} 

This simple question is poorly understood at present. I would like to propose that we {\it use the 'Principle of Domains Correlations' in the multiverse as our criterion for determining the background spacetime(s)} in which the multiverse exists. According to this principle the existence of correlations among domains in the multiverse determines the background spacetime(s) on which the multiverse resides. Thereby the multiverse species fall in two broad distinguishable categories, $connected$ and $disconnected$ as follows:

{\bf Connected:} {\it If different domains in the multiverse are correlated then they must exist in the same background spacetime}.\\

{\bf Disconnected:} {\it If there are no correlations among domains in the multiverse, i.e. domains are totally disconnected from each other, then they must exist in different background spacetimes}.\\

The type $C$ multiverse could provide an illustration of the $Disconnected$ category unless the dimensionally different universes can all be embedded into a higher dimensional space into which correlations are carried via bulk space, as is the case in some brane-world scenarios \cite{randallsundrum} where gravitons spread in the bulk. If this is the case, then we have an example of a multiverse which 'lives' in higher dimensional space where these correlations exist although its member ('brane') universes belong to a lower space dimensionality.

The Connected Multiverse is the most interesting type, not only because it provides one embedding background space and observational handles but also because we have good indications, through CMB, that our universe may be a domain in one such type of the multiverse. Such was the case of the string theory multiverse discussed in \cite{avatars1,avatars2,rudnick}). As wasn shown in \cite{avatars1,avatars2} connectivity through the nonlocal entanglement of our domain with everything else on the multiverse left its imprints on the cosmic microwave background (CMB) and large scale structure (LSS), in various observables. Among them, it is worth mentioning: the prediction of a giant void \cite{avatars1} whose existence was later confirmed experimentally\cite{rudnick}; a suppressed $\sigma_8$ but an enhanced power with distinct signatures at higher multipoles in the power spectrum which is in agreement with the latest WMAP data release\cite{wmap}; and, a higher $SUSY$ breaking scale which will be tested this year by $LHC$. Thus the connected multiverse has the potential for providing observational clues. The Everett and Landscape multiverses, and in general the Type $B$ predictions fall under the 'Connected Multiverse' category.
If time is fundamental then a hierarchy of multiverses is possible with the Disconnected Multiverse type being the embedding space of all different sectors, such that Type $B$ is a sector of Type $A$, and type $B$ and $A$ are subsets of the larger multiverse, type $C$. 

The pressing question related to the variety of multiverse species investigated here is: {\it which one of them is real}? This question stands on shaky foundations since it can be answered only if we could define what constitutes physical reality. Although such definition does not yet exist, we all share the common notion that the existence of spacetime is part of the physical reality and that there can be but one reality. But having only one physical reality implies that only one of the multiverses can correspond to that physical reality. Following this argument seems to place the $Disconnected$ multiverse at a disadvantage. The lack of correlations among domains and the separate existence of sectors in different spacetimes means that there is no way or need for us to ever be aware of the other sectors existence in which case, for all relevance purposes, they are not part of {\it our} physical reality. If nature is economical then there is no need for the $Disconnected$ type which would make their existence unlikely. For the case of $Connected$ universes, due to the unitarity principle, those correlations will continue to exist at all times and all the sectors of this multiverse share the same background spacetime. Therefore all parts of the $Connected$ multiverse are relevant parts of reality for all times, which could make them a better candidate for describing nature.
   
Many of the issues discussed here raise more questions than provide answers. Yet we are on the first steps\cite{tegmark, laurareview, laurarich1,laurarich2} towards addressing a major problem, the ontology of the multiverse, that deserves deeper investigation.

\section{ Discussion}

Extending our physical theories to a multiverse framework may prove to be not only a neccesity for our understanding but also a fertile direction for exploring fundamental questions about or universe and nature. Progress in this field requires that by now we lay a set of principles and ground rules for discussing space and time in the multiverse and, ultimately reduce the class of candidates corresponding to physical reality to only one multiverse.

In this note I have proposed to apply two principles, that to my opinion, are required to discuss issues of space and time in which the classical multiverse is embedded and in which it is  assumed that quantum mechanics is a universal theory. The proposed principles for the multiverse are:

- the principle of {\it 'No Perpetual Motion',} which states that energy can not be created or destroyed in the multiverse; and,\\
- the principle of {\it 'Domains Correlations'} which states that if correlations among domains in the multiverse exists that there is only one corresponding spacetime for all universes.\\

Together these two principles help determine the background spacetime on which the multiverse is embedded.
Further, a possible existence of two phases, the quantum phase and the classical phase for the multiverse allows us to have a coherent picture for the member universes as they go through a quantum to classical transition and a dual picture in quantum Hilbert space and in real spacetime for the multiverse. These phases also allow us to discuss issues such as ergodicity of the phase space for the quantum multiverse and the possibility of Poincarre recurrences that in principle could continually create duplicates of each member universe. As was shown in \cite{laurarich1,laurarich2} an ergodic phase space for the quantum multiverse is highly unlikely due to the out-of-equilibrium dynamics of the degrees of freedom of each domain. But Poincarre recurrences are a consequence of ergodicity, therefore if ergodicity of the phase space is broken then there is no danger that the universe could fluctuate back close enough to its previous state (thereby producing duplicates) no matter how long we wait for these Poincarre recurrences to occur\cite{laurarich2}.

This paper attempts to touch upon some basic issues related to defining the multiverse and its correspondence to physical reality. Although more than $50$ years have elapsed since the first discussion of the 'many worlds' by Everett, we are in the process of resuming the first steps in setting the foundations and the ontology of the multiverse and of this new field in physics. In order to move forward at this relatively early stage in the field, we need to be clear on our set of principles and definitions, hence the need to open the discussion. Who would have though that Nature would lead us to a situation where a deeper understanding of its mysteries, at the smallest and largest scales, would guide the extension of our physical theories towards the realm of the multiverse?

%\maketitle

%\section{Introduction}

%%%%%%%%%%%%%NewIntroduction%%%%%%%%%%%%%%%%%%%%%%%%%%%

\section { Acknowledgements} 
%\begin{theacknowledgments}

LMH was suported in part by DOE grant DE-FG02-06ER41418 and NSF grant PHY-0553312.
%\end{theacknowledgments}

\end{document}